\documentclass{article}
\usepackage[hang,raggedright,tight]{subfigure}
\usepackage{url}
\usepackage{graphicx}
\usepackage{natbib}

\usepackage[ruled]{algorithm2e}
\SetAlFnt{\algofont}
\SetAlCapFnt{\algofont}
\SetAlCapNameFnt{\algofont}
\SetAlCapHSkip{0pt}
\IncMargin{-\parindent}

\title{Sonification of a Network's Self-Organized Criticality}
\author{PAUL VICKERS and CHRIS LAING {Northumbria University}
\\and TOM FAIRFAX {SRM Solutions}\footnote{This work was supported by the United Kingdom's Technology Strategy Board (grant no. BK008B).Author's address: P. Vickers and C. Laing, Northumbria University, Newcastle upon Tyne NE2 1XE, United Kingdom; email: \{paul.vickers, christopher.laing\}@northumbria.ac.uk; Tom Fairfax, SRM Solutions, The Grainger Suite,Dobson House, Regent Centre, Gosforth, Newcastle upon Tyne, NE3 3PF, UK; email: tom.fairfax@srm-solutions.com}}

\begin{document}

\maketitle
\begin{abstract}
Communication networks involve the transmission and reception of 
large volumes of data. Research indicates that network 
traffic volumes will continue to increase. These traffic volumes will 
be unprecedented and the behaviour of global information infrastructures 
when dealing with these data volumes is unknown. It has been tshown 
that complex systems (including computer networks)  exhibit self-organized 
criticality under certain conditions. Given the possibility in such
systems of a sudden and spontaneous system reset the development 
of techniques to inform system administrators of this behaviour could be beneficial.
 This article focuses on the combination of two dissimilar research concepts,
 namely sonification (a form of auditory display) and self-organized criticality 
 (SOC). A system is described that sonifies in real time an information 
 infrastructure's self-organized criticality  to alert the network administrators 
 of both normal and abnormal network traffic and operation.
 \end{abstract}

\section{Introduction}
\label{sec:intro}
With the large volumes of traffic passing across networks
 it is important to know about the state of the various  components involved 
 (servers, routers, switches, firewalls, computers, network-attached storage devices, etc.) 
 and the types and volume of the data traffic passing through the network.
  In the case of the hardware, network administrators need to know if a 
  component has failed or is approaching some capacity threshold (e.g., a 
  server has crashed, a hard drive has become full, etc.) so that appropriate 
  action can be taken. Likewise, the administrators need to be aware of
   traffic type and flow. For example, a large increase in traffic volume 
   (perhaps as would occur if the network were to broadcast a live stream 
   of a major sporting event) might require extra servers to be brought online 
   to handle and balance the load. A sudden increase in certain types of traffic (such as small UDP  packets) might indicate that a distributed denial-of-service attack is in progress, for example, and corrective action would need to be taken to protect the network.\footnote{UDP, or user datagram protocol, is a way of sending internet packets without handshaking. It means that packets can be lost, but in some real-time systems (e.g., online gaming) it is preferable to lose a packet than to wait for a delayed one.}

Given the large volume of traffic passing through a network every second in the form of data packets and the fact that each packet will be associated with particular sender and receiver IP addresses and port numbers, understanding what is happening to a network requires information about the traffic data to be aggregated and presented to the network administrator in an easy-to-understand way. This problem of information presentation and interpretation, or `situational awareness', was addressed by the military leading to Boyd's OODA (observe, orient, decide, act) model (see \citep{Angerman:2004}), and others have followed (notably Endsley's three-level model \citep{Endsley:1995}). Situational awareness, as Cook put it, ``requires that various pieces of information be connected in space and time'' (Nancy Cooke in \citep{McNeese:2012}). 

Computer networks possess high tempo and granularity but with low visibilty and tangibility. Administrators rely on complex data feeds which typically need translatation into language that can be understood by decision makers. Each layer of analytical tools added can increase the margin for error as well as adding Clausewitzian friction (see von Clausewitz's `\emph{On War}', 1873). Furthermore, it is practically impossible for most administrators to watch complex visual data feeds concurrently with other activity without quickly losing effectiveness \citep{Fairfax:2014}.

Thus, in this environment where human perception is constrained, adversaries and protagonists alike are dependent on tools for their perception and understanding of what is going on. Many tools on which we rely for situational awareness are focused on specific detail. The peripheral vision (based on a range of senses) on which our instinctive threat models are based is very narrow when canalised by the tools we use to monitor the network environment. The majority of these tools use primarily visual queues (with the exception of alarms) to communicate situational awareness to operators. Put simply, situational awareness is the means by which protagonists in a particular environment perceive what is going on around them (including hostile, friendly, and environmental events), and understand the implications of these events in sufficient time to take appropriate action.

When network incidents occur experience shows that  the speed and accuracy of the  initial response are critical to a successful resolution of the situation. Operators observe the indicators, orient themselves and their sensors to understand the problem, decide on the action to be taken, and act in a timely and decisive way (OODA).  Traditional approaches to monitoring can hinder this by not making the the initial indication and its context clear thus requiring an extensive orientation stage. An ineffectiveinitial response is consistently seen to be one of the hardest things for people to get right in practice \citep{Fairfax:2014}.
D'Amico (see \citet{McNeese:2012}) put the challenge of designing visualizations for situational awareness this way:
\begin{quotation}
\ldots visualization designers must focus on the specific role of the target user, and the stage of
situational awareness the visualizations are intended to support: perception, comprehension,
or projection.
\end{quotation}

While work has been carried out to use information visualization techniques on network data we note that the \emph{perceive} and \emph{comprehend} stages in Endsley's three-level situational awareness model (the third being \emph{project}) \citep{Endsley:1995} align themselves with Pierre Schaeffer's two fundamental modes of musical listening, \emph{{\'e}couter} (hearing, the auditory equivalent of perception) and \emph{entendre} (literally `understanding', the equivalent of comprehension). \citet{Vickers:2012} demonstrated how Schaeffer's musical context can be applied sonification. This paper proposes a sonification tool as one of the means by which situational awareness in network environments may be facilitated. A more detailed discussion of situational awareness and its relationship to network monitoring (specifically within a cybersecurity and warfighting context) can be found in \citet{Fairfax:2014}.

\subsection{Sonification for Network Monitoring}

Sonification has been applied to many different types of data analysis (for a recent and broad coverage see \emph{The Sonification Handbook} \citep{Hermann:2011}). One task for which it seems particularly well suited is live monitoring, as would be required in situational awareness applications \citep{Vickers:2011}. The approach described in this article provides one way of addressing the challenges outlined above by enabling operators to monitor networks concurrently with other tasks using additional senses.  This has the potential to increase operators' available bandiwidth without overloading individual cognitive functions, and could provide an immediate and elegant route to practical situational awareness.

It has been suggested that understanding the patterns of network traffic is essential to the analysis of a network's survivability \citep{Guo:2008}. Typically, analysis takes place post-hoc through an inspection of log files to determine what caused a crash or other network event. Lessons would be learned and counter measures put in place to prevent a re-occurrence. 

For the purpose of keeping a network running smoothly  load balancing can
sometimes be achieved automatically by the network 
itself, or alerts can be posted to trigger a manual response by the network 
administrators. \citet{Guo:2008} 
observed that ``from the perspective of traffic engineering, understanding 
the network traffic pattern is essential'' for the analysis of network survivability. 

Often, the first the administrators 
know about a problem on a network is after an attack, or other destabilizing event, has 
taken place or the network has crashed. Here, the traffic logs would
 be examined to identify the causes and steps would be taken to try to protect against
 the same events in future. Live 
 monitoring of network traffic assists with situational awareness and could provide administrators either with advanced 
 warning of an impending threat or with real-time intelligence on network 
 threatening events in action.\footnote{By threat, we do not only mean a 
 hacking/DDOS attack, but also include `natural' disasters such as component failures, legitimate traffic surges, etc.} 

Examples of network sonification already exist with Gilfix and Crouch's \textsc{peep} system \citep{Gilfix:2000} being an early example. More recently Balllora et al. \citeyear{Ballora:2011,Ballora:2012} addressed
 situational awareness by letting network behaviour patterns emerge from a soundscape.  They based their approach on the five-level JDL fusion model which aims to enhance situational awareness through the integration of multiple data streams (see \citet{Blasch:2002}). However, they did observe that  the high data speeds and volumes of network traffic could give rise to  unmanageable cognitive loads. Endsley and Connor (in \citet{McNeese:2012}) came to the same conclusion stating that the ``extreme volume of data and the speed at which that data flows rapidly exceeds human cognitive limits and capabilities.'' 
 \citet{Ballora:2011} concluded:
\begin{quote}The combination of the text-based format commonly used in cyber
security systems coupled with the high false alert rates can lead to
analysts being overwhelmed and unable to ferret out real intrusions and
attacks from the deluge of information. The Level 5 fusion process
indicates that the HCI interface should provide access to and human
control at each level of the fusion process, but the question is how
to do so without overwhelming the analyst with the details.
\end{quote}
 
To address managing the complexity we propose that the study of
self-organized criticality has the potential to provide a way of aggregating network behaviour and presenting the `health' of the network as a simple variable, or set of related variables.

\section{Self Organized Criticality in Network Traffic}
Modern networks demonstrate periods of very high activity alternating 
with periods of relative calm, a characteristic known as `burstiness'  \citep{Leland:1993}.
It was commonly thought that ethernet traffic displayed Poisson 
or Markovian distributions. Traffic would thus possess a characteristic burst length which would be smooth when averaged
over a time scale \citep{Crovella:1997}. However, network traffic has been shown to have significant variance or 
burstiness over a range of time scales. Such traffic can be described using the statistical concept
 of self-similarity and it has been established that ethernet traffic
 exhibits this \citep{Valverde:2002}. 
 
 In a wavelet 
analysis of the burstiness of self-similar computer network traffic \citet{Yang:2006} demonstrated  
that the avalanche volume, duration time, and the inter-event time of traffic
 flow fluctuations obey power law distributions.  According to \citet{Bak:1987} 
  such power law distributions in complex systems are evidence of \emph{self-organized criticality} (SOC). 
  SOC is a function of an external
 driving force and and internal relaxation process with  
  a separation of time scales between them \citep{Fairfax:2014}. For example, consider an earthquake. Stresses are the external driving force within 
 the tectonic plates and can take many years to develop. The the internal relaxation process is the earthquake itself and this can
take only seconds, thus there is a separation 
 of time scales between years and seconds \citep{Fairfax:2014}. This separation also 
 comprises two other essential elements: thresholds and metastability \citep{Jensen:1998}. It has been suggested that 
 SOC might be a better explanation of network traffic than traditional models \citep{Yang:2006}. 
  
Since the time taken before an internal relaxation process occurs is non-deterministic, so is the threshold at which 
the internal relaxation process occurs. Thus,  a system can exhibit many many differing states, each of which is
`barely stable', a condition called metastability \citep{Bak:1987}.  \citet{Valverde:2002} showed how network traffic exhibits  the critical states associated with SOC.  

\subsection{Identifying and Measuring the SOC}
\label{sec:measure}

SOC is not a discrete variable that can be identified and monitored directly. Instead, its presence is inferred through the analysis of a system's behaviour or properties, specifically by looking at some time-dependent characteristics. For networks such analysis would typically focus on the traffic, that is, the packets
passing through the system. 
We may observe the SOC by measuring these time-dependent characteristics and comparing changes in successive samples. This is typically done by calculating a log return. The log return, $r$, of two data values on a stream $S$ at intervals $t$ and $t'$ is given by the formula:
\begin{equation}
r = \log[S(t')]-\log[S(t)]\label{logR}
\end{equation}
That is, two successive data samples are converted to logarithms, which are then subtracted to give the log return value. During normal behaviour the log return differences will be small. However, a repeated series of large changes may well indicate a network instability, and the possibility of some form of network `reset'.  Here, a reset does not necessarily mean a catastrophic failure of the network, but could rather  mean the existence of a rapidly increasing level of service traffic restrictions \citep{Fairfax:2014}.

Some simple examples may illustrate what we mean by service traffic restrictions.  The log returns ($r$) of normal network traffic and a network undergoing a distributed denial of service (DDoS) attack were compared using a Daubechies wavelet (part of the wavelet transformation package within Matlab).  Since we are concerned with the notion of self-similar properties, then it made sense to use this particular approach \citep{Brito:1998}.

\begin{figure}[htb]\centering
\subfigure[]{
\includegraphics[width=0.8\linewidth]{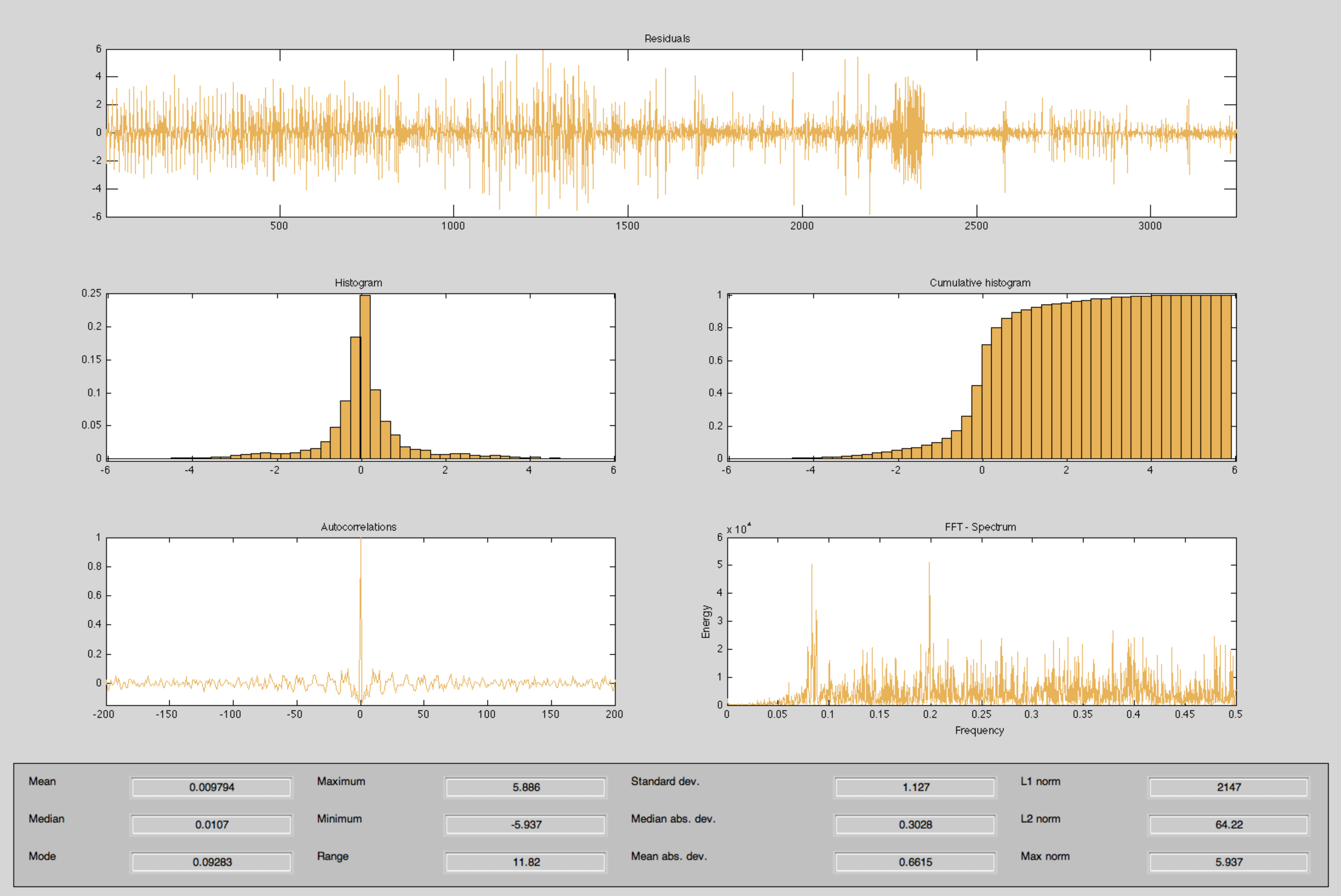}
\label{fig:residuals}}
\subfigure[]{
\includegraphics[width=0.8\linewidth]{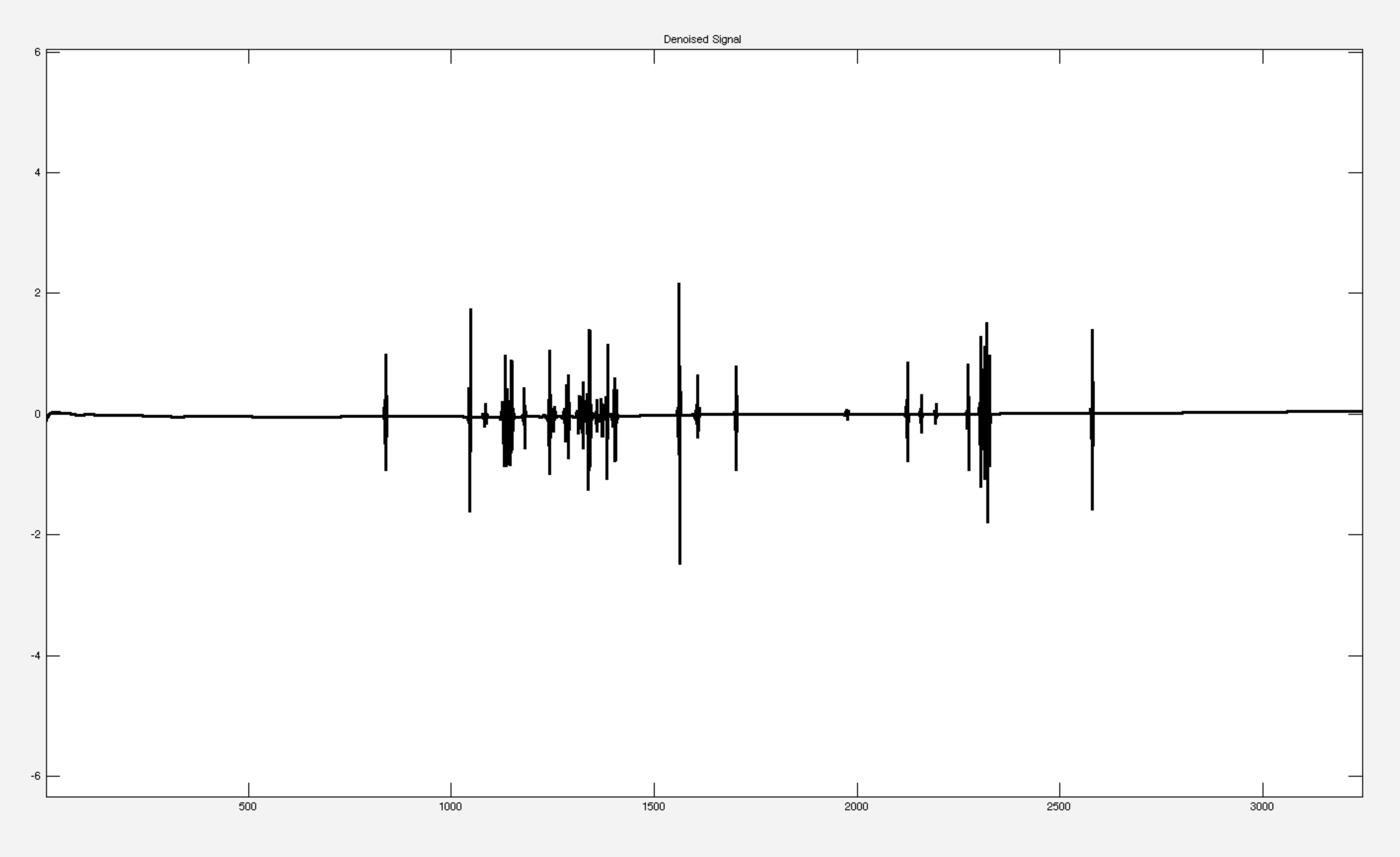}
\label{fig:residualsdenoised}}
\caption{(a) shows a Daubechies wavelet analysis of normal traffic data while (b) shows the denoised traffic residuals.}
\label{fig:normal}
\end{figure}

As can be seen in Figure \ref{fig:residuals}, the residuals have the characteristic burstiness of normal network traffic.  This can be seen more clearly in Figure \ref{fig:residualsdenoised}, were the residuals have been denoised.  In addition, the FFT spectrum for the normal traffic displays almost consistent energy levels across the entire frequency range.

\begin{figure}[htb]\centering
\subfigure[]{
\includegraphics[width=0.8\linewidth]{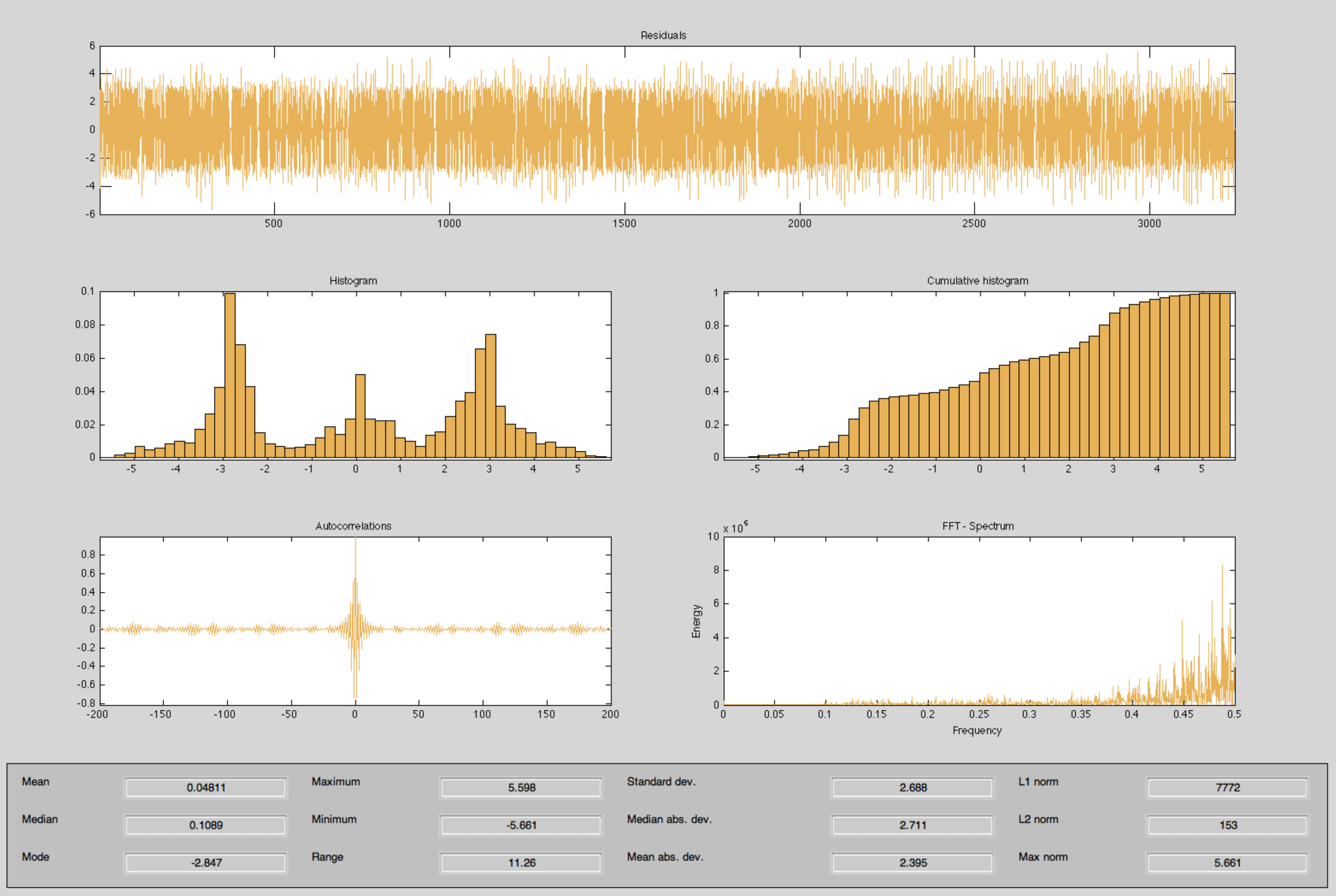}
\label{fig:DDoSresiduals}}
\subfigure[]{
\includegraphics[width=0.8\linewidth]{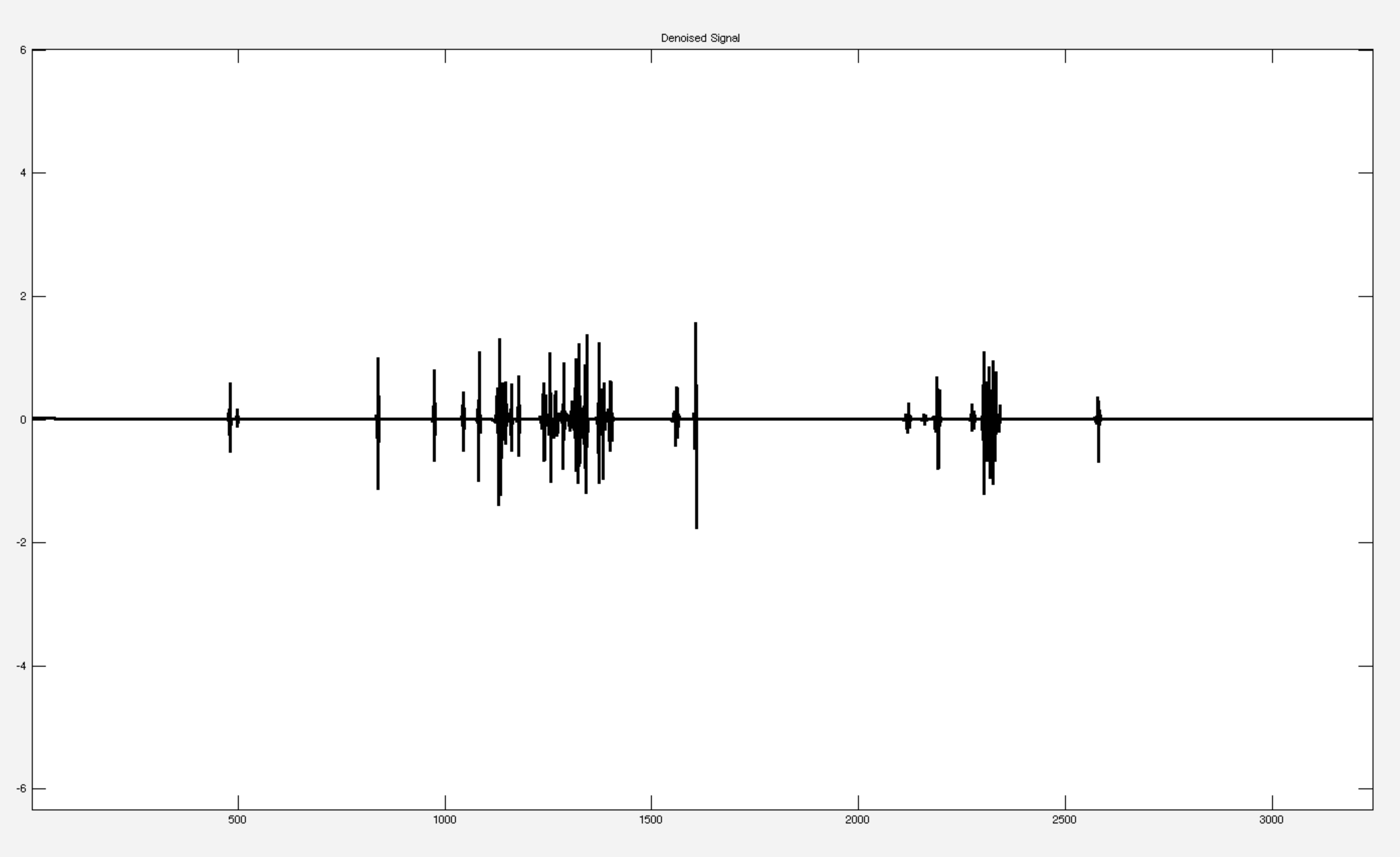}
\label{fig:DDoSresidualsdenoised}}
\caption{(a) Shows a Daubechies wavelet analysis of DDoS traffic data while (b) presents the denoised traffic residuals.}
\end{figure}

Figure \ref{fig:DDoSresiduals} shows DDoS attack traffic.  Again the characteristic burstiness can be seen in the residuals, this time slightly more intense and regular.  However, note the energy levels and distribution in the FFT spectrum.  The energy levels have increased by a factor of 10, while the distribution is confined towards the upper end of the frequency spectrum, and note the rising trend, possibly an indication of increasing SOC activity, and an unstable situation.  

In Figure \ref{fig:DDoSresidualsdenoised}, the residuals of the beginning of a malicious network attack have been denoised. On a cursory inspection it appears to be very similar to Figure \ref{fig:residualsdenoised}. Both figures consist of the same data sets, but Figure \ref{fig:DDoSresidualsdenoised} is a representation of the denoised residuals of normal traffic data that is also carrying malicious traffic data.  Consequently, one would expect to see some differences, and on a closer inspection, the differences become clear.  Firstly, at approximately 500 ($x$-axis) in Figure \ref{fig:DDoSresidualsdenoised}, a small amount of SOC activity can be observed (this is not present in Figure \ref{fig:residualsdenoised}).  Secondly, between 1000 and 1500 ($x$-axis) on both figures, it can be seen that the level and intensity of SOC activity has increased in Figure \ref{fig:DDoSresidualsdenoised}.  Whereas between 1500 and 2000 ($x$-axis), the SOC spike has moved, while between 2000 and 2500 ($x$-axis), the SOC activity has intensified. In the next section we  describe a system for sonifying the SOC characteristics of network traffic.

\section{The SOC Sonification System}
\label{sec:SOCS}

\begin{figure}[htbp]
\begin{center}
\includegraphics[width=\linewidth]{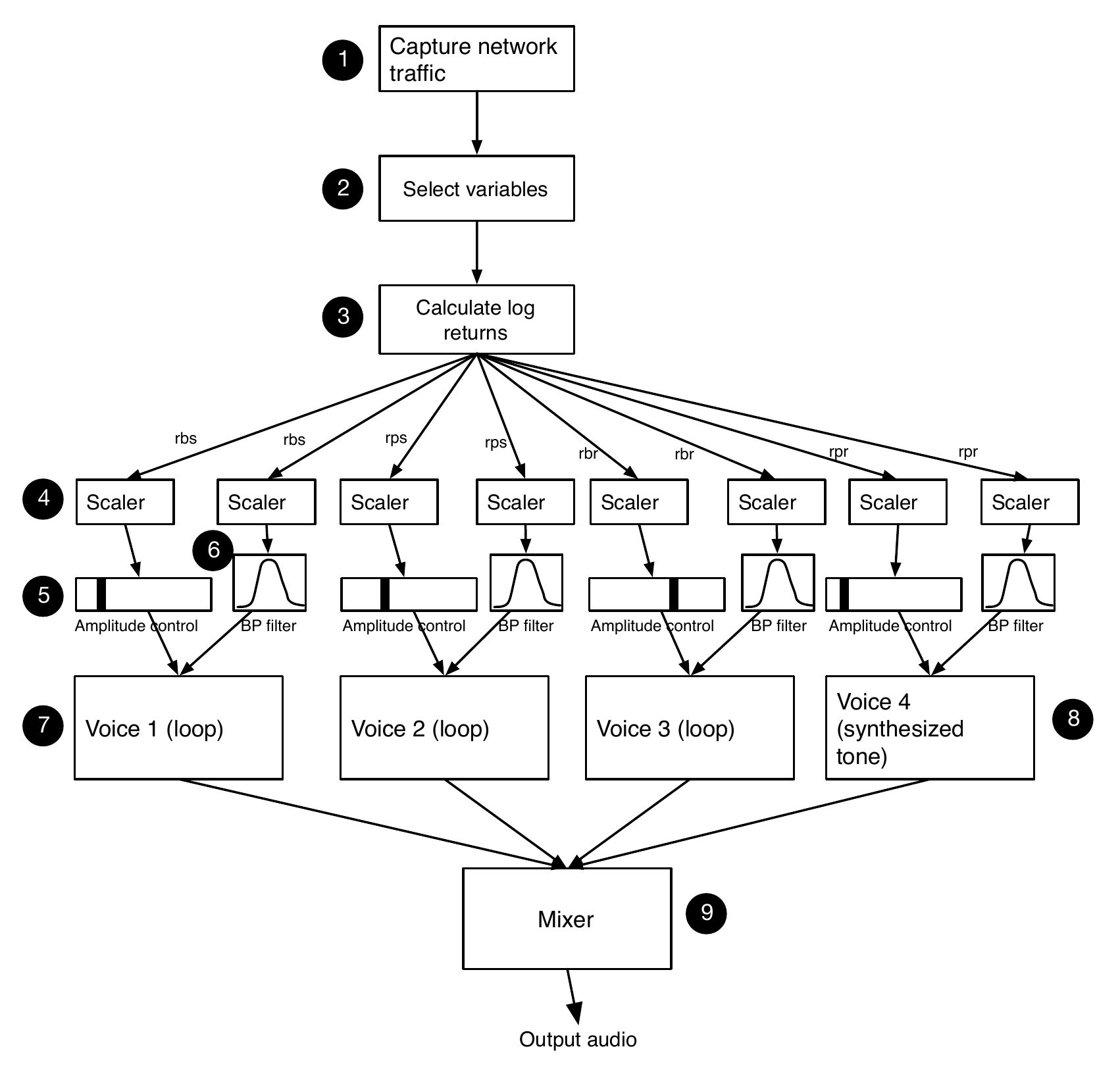}
\caption{Schematic view of the SOC sonification system  for four network traffic 
variables: number of bytes sent ($bs$), number of packets sent ($ps$), 
number of bytes received ($br$), number of packets received ($pr$).}
\label{fig:flowChart}
\end{center}
\end{figure}

Figure \ref{fig:flowChart} shows how SOC sonification system is structured. Network traffic is fed into the 
system either via a live capture device (e.g., the Wireshark program) or 
from a file of previously captured data that is played back via a script to 
maintain the original timing of the events.  For purposes of illustration, the example 
chosen here sonifies the log returns of the following time-dependent network traffic data items: 
number of bytes sent, number of packets sent, number of bytes received, 
number of packets received by the network which we call $bs$, $ps$, $br$, $pr$ 
respectively. These variables represent the total number of packets and bytes sent 
and received in a given time interval, $t,t'$. As SOC has been shown to exist across 
multiple timescales, network traffic could be sampled at any regular interval. 
The size of the interval is not specified and is at the discretion of the user. 
SOC properties can be observed by comparing the log return values of
successive samples of time series data.  
Thus, in 
this example we calculate four log return values for the variables 
$bs$, $ps$, $br$, $pr$:
\begin{eqnarray}
rbs &=& \ln\left[bs(t')\right]-\ln\left[bs(t)\right]\label{rbs} \\
rps &=& \ln\left[ps(t')\right]-\ln\left[ps(t)\right]\label{rps} \\
rbr &=& \ln\left[br(t')\right]-\ln\left[br(t)\right]\label{rbr}\\
rpr &=& \ln\left[pr(t')\right]-\ln\left[pr(t)\right]\label{rpr} 
\end{eqnarray}
This may result in negative values for the log return which can be 
used to indicate the direction of a SOC event's change in level 
(i.e., an increase in value means a step up to the next level of 
steady state, whilst a decrease means a step down). Alternatively 
the system can also use squared log return values to keep all 
values positive (which might be done if one were interested only 
in large changes of level regardless of direction).

The log return values are fed as input into the sonification tool. 
Each log return value is used to control the parameters of an 
individual sound generator (or \emph{voice}). In the example 
shown here there are four voices but the system can be
extended to include as many voices as there are data dimensions to be monitored. 
A voice can be a synthesized tone generated in real time or it 
can be a segment of sampled audio that is played back as a 
repeated loop. Figure \ref{fig:flowChart} shows a setup 
in which three traffic variables are mapped to loop-based 
voices whilst a fourth is mapped to a synthesized voice. The loops 
and synthesized voices could, in principle, be any sound, but 
it is recommended to use sounds with large broadband noise 
components to minimize perceptual distraction. In the version 
described here, the channels contained different sounds
that combined to make a countryside soundscape (e.g., running stream,
rain, crickets, and so forth). This enables the various audio channels
to be attended to as a single coherent whole, but alterations in any
single channel will stand out.

The log return value of each data stream is used to modulate the 
corresponding voice. This may be done by increasing/decreasing 
the amplitude, altering the voice's position in a sound field (e.g., 
left-right pan in a stereo field, front/back/left/right in a surround-sound 
field, or front/back/left/right and azimuth in a full three-dimensional 
sound field), altering the voice's phase, or altering its spectral 
characteristics (e.g., by changing the parameters of a bandpass filter). 
The example sonification in Figure \ref{fig:flowChart} shows the four 
data streams modulating 1) the amplitude of the voices and 2) 
the centre frequency and resonance value of a bandpass filter. 
Each log return value is scaled to a range of values appropriate 
for the control it is being used to modulate.

\subsection{Explanation of Figure \ref{fig:flowChart}}
Each of the numbered items in Figure \ref{fig:flowChart} is explained below.
\begin{enumerate}
  \item \textsc{Capture network traffic}: Network traffic is captured 
  using a packet sniffer program. This could either be a standard 
  off-the-shelf product (e.g., Wireshark) but in our example we 
  wrote a custom Python script that used the Python \texttt{socket} library.
  \item\textsc{Select variables}: SOC is a property of the network 
  traffic. It can be represented by selecting certain traffic variables 
  and calculating the log returns of successive values of those 
  variables. It may also be possible to measure SOC via a more 
  sophisticated statistical analysis of several variables in combination. 
  To demonstrate the principle of sonifying SOC we chose to 
  measure four variables (number of packets received, number of 
  packets sent, number of bytes received, number of bytes sent) 
  per 1-second interval. These values are extracted from the 
  network traffic data gleaned by the packet sniffer using a 
  Python script.
  \item\textsc{Calculate the log returns}: SOC is evidenced 
  through the rise and fall of log return values of the variables 
  being measured. Calculation of the log returns is done in 
  our Python script. 
  \item\textsc{Scaler}: SOC evidences itself through 
  orders-of-magnitude changes in the log return values. 
  However, audio processing units tend to require restricted 
  ranges of digital input values (say, $0\ldots127$, $0\ldots15$, $-256 \ldots 255$, etc.) 
  Therefore, we built a scaler function that takes four arguments: 
  the minimum and maximum values of the input range and 
  the minimum and maximum of the desired output range. 
  It then scales any value received on its input channel to 
  a corresponding value in the specified output range. 
  \item\textsc{Amplitude control}: There are many possible 
  mappings between the input data values and the various 
  parameters that affect the audio. For example, one could 
  map log return value to frequency or amplitude, or one 
  could use the value to change the behaviour of a filter that 
  would, in turn, alter the harmonic spectrum of the output 
  sound. To show the sonification in practice we used the 
  four tracked variables to control in real-time the amplitude 
  and harmonic spectrum of the audio. The \textsc{amplitude control} 
  is a component that adjusts the amplitude, or level, of 
  the output sound according to the value of the input variable. 
  The lower the log return value the quieter the sound, the 
  higher it is the louder the sound it played back. Thus, the 
  real-time monitoring of the network leads to constant 
  fluctuations in the amplitude of the output (but only 
  large changes in level are readily perceived).
  \item\textsc{BP filter}: There are several ways that the 
  spectral characteristics of an audio signal may be 
  processed, each of which will cause a change in the 
  timbre of the audio. For this example we used the input 
  variable values to determine the central frequency 
  (or `Q' value) of a band pass filter. A band pass filter is 
  a device that allows frequencies within a certain range 
  (centred around `Q') to pass unhindered and which 
  attenuates frequencies falling outside this range. We 
  set up four BP filters (one per audio channel) and manually 
  specified their frequency ranges. The value of the input 
  variable being monitored by each channel was then 
  used to change the Q value of its BP filter in real time. 
  This means that the timbre of the audio changed as 
  the input variable changed. When the values were low, 
  the Q value was set low; high values increased the Q value. 
  This gives the effect of a more muffled sound when the log 
  return values are low and a brighter sound when they are high.
  \item\textsc{Voice $n$ loop}: There are many ways to 
  represent data values in audio. Two ways demonstrated 
  in this example are the looped playback of pre-recorded 
  audio and the generation of a synthesized tone. The 
  \textsc{Voice $n$ loop} module loads a pre-defined audio
   loop and plays it continuously, restarting it when it reaches 
   the end. The amplitude of the loop is controlled by the 
   \textsc{Amplitude control} (above) and its timbre is controlled 
   by a \textsc{BP filter} above.
  \item\textsc{Voice $n$ (synthesiszed tone)}: Any synthesis 
  technique could be used to generate a synthetic tone, but 
  in this example we used a simple a white-noise generator 
  which is filtered and processed to make a synthetic rain sound. 
  The amplitude and timbre of the sound are controlled using 
  the controls discussed above.
  \item\textsc{Mixer}: The four audio channels (one for 
  each of the four variables being monitored) are sent 
  to an audio mixer control which mixes the sounds 
  into a single stereo output which is then sent to the 
  audio system of the host computer.
  \end{enumerate}
 
\begin{figure}[htbp]
  \centering
  \includegraphics[width=\linewidth]{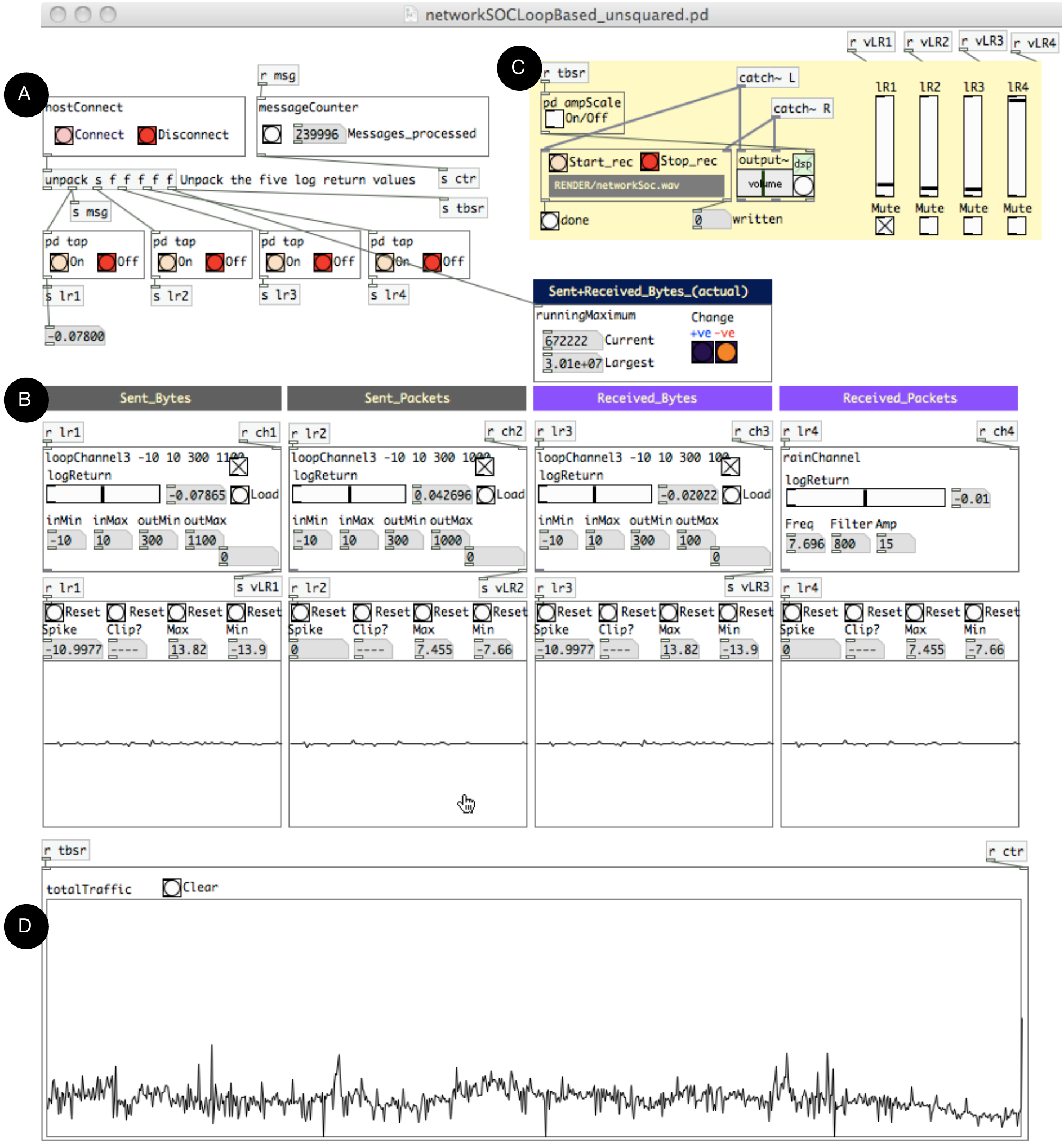}
  \caption{The \textsc{socs} application. Section A deals with reading the 
  network traffic from the capture device. Section B contains the voice
  definitions to which each traffic variable is mapped. Section C is a mixer
  to convert the four separate audio streams into a single stereo feed. Section D
  is a graphical display of the combined variables being monitored. The channel
  graph plots are updated more frequently than the aggregate graph plot.}
  \label{fig:screenShot}
\end{figure}

 The \textsc{socs} (self-organized criticality sonification) 
 tool was implemented using the Pure Data audio 
 programming environment (freely available from \url{http://puredata.info}) and a Python script for 
 dealing with the capture of network packets and 
 the transmission to the tool of the 
 log return values of the variables being monitored. 
 Figure \ref{fig:screenShot} shows a screen shot of the 
 application as it looks to the user.
The application has four principal sections: network input (A), 
channel processing (B), the mixer (C), and the graph view (D). 
The network input section contains a module that receives the log 
returns generated by the Python script. The channel processing 
section contains four similar units: three for dealing with audio loop 
playback and one for dealing with synthesized tone playback. Each 
of the four units contains a scaler module and a band pass filter 
module. The three loop-based units also contain modules for loading 
and playing back the pre-recorded audio files. The synthesizer unit 
contains modules for generating and filtering white noise. Each of 
these four channel processors contain a real-time graphic plot 
which shows the values of the log returns. The mixer section 
(C) allows the relative amplitudes of the four channels to be set. 
These four channels are then mixed down to a single stereo output 
which is sent to the host computer's audio hardware. The graph view 
(D) plots the aggregate network traffic in real-time which allows 
visual reference to be made when something of interest is heard.

The network input section (A) contains taps to turn on and off the four
data streams that are being sonified. This allows the operator to generate 
an overall soundscape of all the network variables being monitored or to focus on 
an desired subset. Additionally, the mixer section (C) allows the overall balance between the 
soundscape channels to be adjusted as desired.

\section{Discussion}
The system was tested with a number of traffic data sets captured from live networks.
Traffic data were aggregated over 1s intervals and the number of bytes and packets sent and 
received per interval were fed to the \textsc{socs} application. Each time a set of log return
values is received the system uses the values to modulate the four respective audio channels.

\begin{figure}[htbp]
  \centering
  \includegraphics[width=\linewidth]{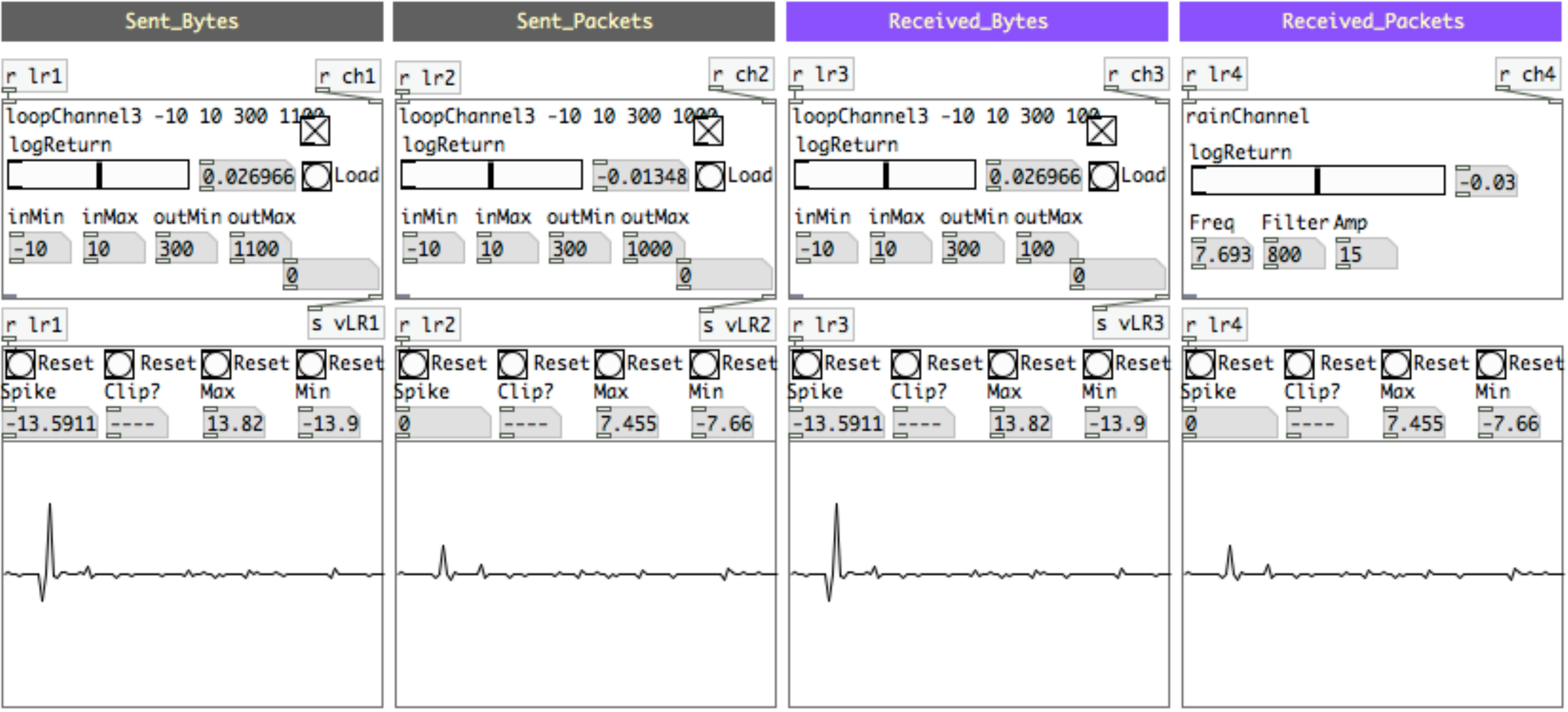}
  \caption{This screen shot of the voice channel section shows
  log return spikes occurring on all four channels with the largest
  values occuring in the sent bytes and sent packets streams. These
  spikes generate a noticeable increase in the amplitude and brightening
  of the timbre of the soundscape.}
  \label{fig:logReturnSpike}
\end{figure}

When the traffic is exhibiting normal patterns small fluctuations in log return values 
do not lead to very noticeable changes in the soundscape, either in amplitude or
timbre. However, when one or more very large log returns occur (such as would be
expected during a dynamic system relaxation event) the corresponding soundscape
experiences a very noticeable change: the amplitude increases greatly and the timbre
becomes brighter as the Q point of the band pass filters is moved up in the frequency range
(see Figure \ref{fig:logReturnSpike}).

On hearing an event like this (situational awareness level 1--- Perception) the network aministrator would be drawn to inspecting the
state of the network (situational awareness level 2 --- Comprehension) to see if any action needs to be taken (situational awareness level 3 --- Projection). Then comes the stage of managing the action, which itself requires situational awareness as actions are taken to address the situation.
The final step in all UK military decision support methodology is to ask the question ``has the situation changed?'', thus restarting the OODA loop.
In a healthy network one 
would expect a number of significant changes in the soundscape over time as relaxation
events occur (much as a sandpile would undergo shifts in its topology as sand is added to
it over time). Some of these events may go unnoticed by the administrator (if, for example,
they left the monitoring station for a short period of time) but this would not be
harmful. What will be of particular interest is when there is an extended series of
repeated high log return values which might indicate growing instability in the network.
An extended period of increased soundscape amplitude signals as a clear alert
to the administrator.

The system was tested with a log return intervals of 1s, but the self-similarity of network
traffic burstiness means that a range of timescales can be used.  Another way the system
was run was to take a log of sampled traffic data and to feed it into the \textsc{socs} system
at a higher rate than its initial sample rate. This allows historic feeds to be listened to post-hoc
in a manner analogous to spooling quickly through  an audio tape (the main difference being
that there is no consequent alteration of pitch).  This means that logs can be auditioned quickly
and interesting areas of activity spotted. 

\section{Concluding Remarks}
The combination of using a system's self-organized criticality as the underlying 
data set for situational awareness and a tool for sonifying this SOC offers a number of 
potential advantages. First, because SOC is an emergent property of the network as a whole, and
can be seen at different time scales, it means that one can get an impression of the overall state
of a network by monitoring a relatively small number of data streams, thereby ameliorating the problems of extreme volumes and speeds of data identified by previous researchers. Second, the sonification approach allows for the real-time presentation of simple, but relevant data via a medium that lets network administrators work at situational awareness levels 1 and 2 using without having to keep a visual focus on a complex graphical display. Third, because SOC manifests itself fractally and across timescales, whatever data sampling interval is chosen, any changes in network criticality will still be identifiable.  

While the work described here focused on the traditional traffic metrics of
bytes and packets sent and received, it will be instructive to explore
what other variables and characteristics are implicated in a network's SOC and this is the subject of ongoing work. 

The present prototype system allowed the creation of a soundscape of up to four independent audio streams (mixed down to
a pair of stereo channels). The underlying system architecture promotes interactivity by letting the user select the combination
of incoming data streams to be sonified and the sonic balance of the auditory streams. Another aspect of the ongoing work
is to explore combining sonification with a multitouch display to create a richer interaction experience. The following example use case describes how this might be realized. A possible intrusion is detected through an anomalous change in the SOC variables. The administrator now wishes to investigate the network’s behaviour. To do this a diagram showing a network setup is projected onto a multi-touch display. The data indicate a problem between the router and the internet, and between the switch and the laptop. A tangible user interface object (e.g., a cube) with a fiducial marker on its bottom surface is placed above the router.1 Another object is placed on the channel between the switch and the laptop. A camera beneath the display recognizes the fiducials which are coded to specific traffic data variables. Rotating the objects controls the auditory and/or visual parameters of the data streams. Visual feedback can be projected onto the surface (e.g., printing data above the interface object) with auditory feedback being via loudspeakers or headphones. The interface objects become probes to monitor chosen network locations for particular events or data types. In this way the administrator can gain intelligence about the state of the network in a hands-on way.

\section*{Acknowledgements}The authors gratefully acknowledge the input of Jonathan Christison, a final-year student on Northumbria University's
BSc Ethical Hacking for Computer Security who provided assistance with constructing the Python packet sniffer.
Patent Applied For: This article was the subject of UK Patent Application no. GB1205564.6 filed on 29 March 2012.
\bibliographystyle{acmlarge}
\bibliography{bibliography}

\providecommand{\noopsort}[1]{}
\begin{thebibliography}{}

\bibitem[\protect\citeauthoryear{Angerman}{Angerman}{2004}]{Angerman:2004}
{\sc Angerman, W.~S.} 2004.
\newblock Coming full circle with {B}oyd's {OODA} loop ideas: An analysis of
  innovation diffusion and evolution.
\newblock M.S.\ thesis, Airforce Institute of Technology, Wright-Patterson AFB,
  Ohio, USA.

\bibitem[\protect\citeauthoryear{Bak, Tang, and Wiesenfeld}{Bak
  et~al\mbox{.}}{1987}]{Bak:1987}
{\sc Bak, P.}, {\sc Tang, C.}, {\sc and} {\sc Wiesenfeld, K.} 1987.
\newblock Self-organized criticality: An explanation of the $1/f$ noise.
\newblock {\em Phys. Rev. Lett.\/}~{\em 59,\/}~4, 381--384.

\bibitem[\protect\citeauthoryear{Ballora, Giacobe, and Hall}{Ballora
  et~al\mbox{.}}{2011}]{Ballora:2011}
{\sc Ballora, M.}, {\sc Giacobe, N.~A.}, {\sc and} {\sc Hall, D.~L.} 2011.
\newblock Songs of cyberspace: An update on sonifications of network traffic to
  support situational awareness.
\newblock {\em Proc. SPIE\/}~{\em 8064}, 80640P--80640P--6.

\bibitem[\protect\citeauthoryear{Ballora, Giacobe, McNeese, and Hall}{Ballora
  et~al\mbox{.}}{2012}]{Ballora:2012}
{\sc Ballora, M.}, {\sc Giacobe, N.~A.}, {\sc McNeese, M.}, {\sc and} {\sc
  Hall, D.~L.} 2012.
\newblock Information data fusion and computer network defense.
\newblock In {\em Situational Awareness in Computer Network Defense:
  Principles, Methods and Applications}, {C.~Onwubiko} {and} {T.~Owens}, Eds.
  IGI Global.

\bibitem[\protect\citeauthoryear{Blasch and Plano}{Blasch and
  Plano}{2002}]{Blasch:2002}
{\sc Blasch, E.~P.} {\sc and} {\sc Plano, S.} 2002.
\newblock {JDL} level 5 fusion model: User refinement issues and applications
  in group tracking.
\newblock In {\em Proc. {SPIE}}. Vol. 4729. 270--279.

\bibitem[\protect\citeauthoryear{Brito, Souza, and Pires}{Brito
  et~al\mbox{.}}{1998}]{Brito:1998}
{\sc Brito, N. S.~D.}, {\sc Souza, B.~A.}, {\sc and} {\sc Pires, F. A.~C.}
  1998.
\newblock Daubechies wavelets in quality of electrical power.
\newblock In {\em Harmonics and Quality of Power Proceedings, 1998.
  Proceedings. 8th International Conference On}. Vol.~1. Athens, 511--515.

\bibitem[\protect\citeauthoryear{Crovella and Bestavros}{Crovella and
  Bestavros}{1997}]{Crovella:1997}
{\sc Crovella, M.~E.} {\sc and} {\sc Bestavros, A.} 1997.
\newblock Self-similarity in world wide web traffic: Evidence and possible
  causes.
\newblock {\em IEEE/ACM Trans. Netw.\/}~{\em 5,\/}~6, 835--846.

\bibitem[\protect\citeauthoryear{Endsley}{Endsley}{1995}]{Endsley:1995}
{\sc Endsley, M.} 1995.
\newblock Toward a theory of situation awareness in dynamic systems.
\newblock {\em Human Factors\/}~{\em 37,\/}~1, 32--64.

\bibitem[\protect\citeauthoryear{Fairfax, Laing, and Vickers}{Fairfax
  et~al\mbox{.}}{2014}]{Fairfax:2014}
{\sc Fairfax, T.}, {\sc Laing, C.}, {\sc and} {\sc Vickers, P.} 2014.
\newblock Network situational awareness: Sonification \& visualization in the
  cyber battlespace.
\newblock In {\em Handbook of Research on Digital Crime, Cyberspace Security,
  and Information Assurance}, {M.~M. Cruz-Cunha} {and} {I.~M. Portela}, Eds.
  Advances in Digital Crime, Forensics, and Cyber Terrorism. IGI Global, in
  press.

\bibitem[\protect\citeauthoryear{Gilfix and Couch}{Gilfix and
  Couch}{2000}]{Gilfix:2000}
{\sc Gilfix, M.} {\sc and} {\sc Couch, A.~L.} 2000.
\newblock Peep (the network auralizer): Monitoring your network with sound.
\newblock In {\em 14th System Administration Conference ({LISA} 2000)}. The
  {USENIX} Association, New Orleans, Louisiana, USA, 109--117.

\bibitem[\protect\citeauthoryear{Guo, Wang, nv~Huang, and Zhao}{Guo
  et~al\mbox{.}}{2008}]{Guo:2008}
{\sc Guo, C.}, {\sc Wang, L.}, {\sc nv~Huang, L.}, {\sc and} {\sc Zhao, L.}
  2008.
\newblock Study on the internet behavior's activity oriented to network
  survivability.
\newblock In {\em International Conference on Computational Intelligence and
  Security, 2008. {CIS} `08}. Vol.~1. {IEEE}, 432--435.

\bibitem[\protect\citeauthoryear{Hermann, Hunt, and Neuhoff}{Hermann
  et~al\mbox{.}}{2011}]{Hermann:2011}
{\sc Hermann, T.}, {\sc Hunt, A.~D.}, {\sc and} {\sc Neuhoff, J.}, Eds. 2011.
\newblock {\em The Sonification Handbook}.
\newblock Logos Verlag, Berlin.

\bibitem[\protect\citeauthoryear{Jensen}{Jensen}{1998}]{Jensen:1998}
{\sc Jensen, H.~J.} 1998.
\newblock {\em Self-Organized Criticality}.
\newblock Cambridge University Press, Cambridge.

\bibitem[\protect\citeauthoryear{Leland, Taqqu, Willinger, and Wilson}{Leland
  et~al\mbox{.}}{1993}]{Leland:1993}
{\sc Leland, W.~E.}, {\sc Taqqu, M.~S.}, {\sc Willinger, W.}, {\sc and} {\sc
  Wilson, D.~V.} 1993.
\newblock On the self-similar nature of {Ethernet} traffic.
\newblock {\em SIGCOMM Comput. Commun. Rev.\/}~{\em 23,\/}~4, 183--193.

\bibitem[\protect\citeauthoryear{McNeese}{McNeese}{2012}]{McNeese:2012}
{\sc McNeese, M.} 2012.
\newblock Perspectives on the role of cognition in cyber security.
\newblock In {\em Proceedings of the Human Factors and Ergonomics Society 56th
  Annual Meeting -- 2012}. 268--271.

\bibitem[\protect\citeauthoryear{Valverde and Sol{\'e}}{Valverde and
  Sol{\'e}}{2002}]{Valverde:2002}
{\sc Valverde, S.} {\sc and} {\sc Sol{\'e}, R.~V.} 2002.
\newblock Self-organized critical traffic in parallel computer networks.
\newblock {\em Physica A\/}~{\em 312}, 636--648.

\bibitem[\protect\citeauthoryear{Vickers}{Vickers}{2011}]{Vickers:2011}
{\sc Vickers, P.} 2011.
\newblock Sonification for process monitoring.
\newblock In {\em The Sonification Handbook}, {T.~Hermann}, {A.~D. Hunt}, {and}
  {J.~Neuhoff}, Eds. Logos Verlag, Berlin, 455--492.

\bibitem[\protect\citeauthoryear{Vickers}{Vickers}{2012}]{Vickers:2012}
{\sc Vickers, P.} 2012.
\newblock Ways of listening and modes of being: Electroacoustic auditory
  display.
\newblock {\em Journal of Sonic Studies\/}~{\em 2,\/}~1.

\bibitem[\protect\citeauthoryear{Yang, Jiang, Zhou, Wang, and Zhou}{Yang
  et~al\mbox{.}}{2006}]{Yang:2006}
{\sc Yang, C.-X.}, {\sc Jiang, S.-M.}, {\sc Zhou, T.}, {\sc Wang, B.-H.}, {\sc
  and} {\sc Zhou, P.-L.} 2006.
\newblock Self-organized criticality of computer network traffic.
\newblock In {\em Communications, Circuits and Systems Proceedings, 2006
  International Conference on}. Vol.~3. {IEEE}, 1740--1743.

\end{thebibliography}
\end{document}